\documentclass[12pt]{article}
\usepackage[utf8]{inputenc}
\usepackage{amsmath,amsthm,amsfonts,amssymb}

\begin{document}

\title{Caustics type of gravitational singularities}

\author{G. A. Sardanashvily,\and E. G. Timoshenko\thanks{Translated on 07.07.2024 by Edward.Timoshenko@ucd.ie from Russian: The Bulletin of Moscow State University, ISSN 2074-6636, 
UDC 530.12, Series 3, Physics and astronomy, Vol. 30, No 3, pp. 75-77 (1989).}
}

\date{
Division of Theoretical Physics, Department of Physics, Lomonosov Moscow State University, Russia, USSR
}

\maketitle

\begin{abstract}
We describe a new type of gravitational singularities which are caustics of spatial--temporal foliations.
\end{abstract}


\section{Introduction}
\label{sec:intro}

The description of gravitational singularities as singularities of space--time foliations [1-3] allows one, unlike other methods, to establish the structure of these singularities. In this paper, we consider the caustics of spatial--temporal foliations.

There is a diagram
\[
\begin{array}{ccc}
GL(4, \mathbb{R}) & \longrightarrow & SO(4) \\
\downarrow & & \downarrow \\
SO(3,1) & \longrightarrow & SO(3)
\end{array}
\]
of the reduction of the structural group of the tangent bundle $T(X^4)$ over a smooth orientable paracompact manifold $X^4$ admitting the gravitational field $g$ as a global section of the bundle of pseudo--Euclidean bilinear forms in tangent spaces to $X^4$. By virtue of this diagram, any gravitational field on the manifold $X^4$ defines some smooth orientable space--time distribution $F$ of 3-dimensional space--like (with respect to $g$) hyperplanes of tangent spaces to $X^4$, given by the equation $h^0(F) =0$, where $h^0=h^0_{\mu}dx^{\mu}$ is the tetrad form of the field $g$. Inversely, any 3-dimensional smooth orientable distribution on $X^4$ is a space--time with respect to some gravitational field [2,3].

A distribution $F$ is called integrable (foliation) if its generating form $h^0$ satisfies the equation $h^0 \wedge dh^0=0$. In this case, the manifold $X^4$ splits into 3-dimensional space--like hypersurfaces such that the layers of the distribution are tangent spaces to these hypersurfaces.
A space--time foliation is said to be causal if $h^0=Ndf$, where $N$ and $f$ are real functions on $X^4$ such that $N\not= 0$ and $df\not= 0$ are everywhere on $X^4$.

The correspondence between gravitational fields and spacetime distributions on the manifold $X^4$ allows us to describe gravitational singularities as singularities of such distributions. Since any regular gravitational field locally admits a causal foliation, singularities of distributions can be described as singularities of causal foliations in terms of generating functions $f$.
Two types of such singularities are distinguished.

1. The generating function $f$ is single--valued but has a critical point where $df=0$. It gives rise to a Hefliger structure (singular foliation of  $f$ level surfaces on $X^4$, which change their topology when passing through the critical point of the function $f$. 

2. The function $f$ is multivalued. In the region where it is single-valued, the function $f$ generates a foliation, which, however, collapses at branching points 1, where the layers begin to intersect each other. To describe this, we can lift the foliation $F$ into the total space of the cotangent bundle $\mbox{tl}\ T^{*}(X^4)$, continue it there and then project it back to the base $X^4$. The singularities of $F$ will correspond to the singularities of this projection, which are caustic points.

In gravitation theory (by analogy with geometrical optics), a caustic is usually a geometric place of focal or conjugate points [4,5]. We follow a more general definition of caustics as singularities of Lagrangian mappings [6]. Any such caustics can be locally (in terms of germs) reduced to the following canonical construction.

Consider space $R^{2n}$ with coordinates $(x^{\mu},P_{\mu})$ 1- form $a=P_{\mu}dx^{\mu}$ on $R^{2n}$ and a $n$-dimensional submanifold $N \subset R^{2n}$ such that $da|_N=0$. This is called Lagrangian submanifold and 
it can be defined by the generating function $S(x^i, P^j)$ of $n$ variables $(x^i, P_j,\ i \in I,  j \in J)$, where $(I, J)$ is some partition of the set ($1$,\dots $n$), and is defined by the relations
\[
x^j = -\frac{\partial S}{\partial P_j}, \quad P_i = \frac{\partial S}{\partial x^i}.
\]

Let $\pi: \ \{x^{\mu}, P_{\mu}\} \to  \{x^{\mu}\}$ be the projection of $R^{2n}$ onto $R^n$. Being restricted on the Lagrangian manifold:
\[
\pi_N : (x^i, P_j) \rightarrow (x^i, x^j = -\partial S / \partial P_j),
\]
it is called a Lagrangian mapping, and the set of its special points (where the matrix $\partial^2 S/\partial P_i \partial P_j $ is degenerate) is called a caustic.

Let $f$ be the generating function of the foliation $F$ on $X$. Let us define the embedding
\[
\gamma: \{x^\mu\} \to \{x^\mu, P_\mu = \partial f / \partial x^\mu\}
\]
of manifold $X$ into the total space tl$\,T^*(X)$. Its image is a Lagrangian submanifold of tl$\,T^*(X)$ endowed with an induced foliation $F'= \pi^*_{\gamma(X)}F$, where $\pi_{\gamma(X)}:\ \gamma(X)\to X$ is a Lagrangian mapping. Inversely, let $N \subset \mbox{tl}\, T^*(X)$ be a Lagrangian submanifold with a generating function $S(x^i, P_j)$. Let $F'$ denote the foliation of level surfaces of the function
\[
f'(x^i, P_j) = S - P_j \frac{\partial S}{\partial P_j}
\]
on $N$. The image $\pi_N(F')$ of the foliation $F$ under the Lagrangian mapping  $\pi_N$ forms some structure on $X$, which is a foliation with the generating function $f(x^{\mu}) = f'(x^i, P_j(x^i, x^j))$ on the region where $\pi_N$  has no singularities. This foliation breaks down at the caustic points of the mapping $\pi_N$, where the functions $P_j(x^i, x^j)$ and $f(x^{\mu})$ become multivalued. The classification of stable caustics $A_{2-5},\ D_{4,5}$ is known on a 4-dimensional manifold.

Gravitational singularities of the caustics type have the following distinctive feature. There exist regions where not the nearest, but, on the contrary, the layers of foliation far separated by the values of $f$ intersect. Therefore, the spatial--temporal foliation and, consequently, the gravitational field in such a region can be locally extended beyond the caustic points. Let us give the following example.

We define $f(u,v)$ as a function on $R^2$ satisfying the equation
$f^3-3uf-2v=0$. This function is single-valued:
\[
f_+ = [v + (v^2 - u^3)^{1/2}]^{1/3} + [v - (v^2 - u^3)^{1/2}]^{1/3}
\]
in the region $U= (u,v: v^2>v^3)$ and is three--valued:
\[
f_{0,1,2} = 2u^{1/2} \cos\left(\frac{1}{3}(\varphi + 2\pi n)\right), \quad \varphi = \arccos(vu^{-3/2}), \quad n = 0, 1, 2
\]
at points $v^2<u^3$. Let $F$ be a foliation
$f_+(u, v)=c=const$ on $U\subset R^2$. 
Its layers $F_c$ 
are straight lines $2v=c^3-3 cu$. This foliation forms a caustic at the branching points
$v^2=u^3$, so that $u=v=0$ is a caustic of type $A_3$ whereas $v^2=u^3 \not= 0$
$A_2$-type caustic points. The layers $F_c$, 
$c>\varepsilon >0$ can be extended beyond the caustic line $v=u^{3/2}$ as $f_0(u,v)=const$ layers in the region $0<v<u^{3/2}$. But they begin to intersect each other when $v<0$, although the layers closest to each other intersect only near the caustic line $v =-u^{3/2}$. At the same time, the layers $F_{c>0}$ begin to intersect with the layers $F_{c<0}$ already at the caustic line 
$v = u^{3/2}$. 

An example of singularity with caustics of the above type gives the gravitational wave $(h^0=Nd(f_+ + x^2))$
\[
ds^2 = 2dx^2du + \left[-f_{+,v}^2-(f_{+,u}-1)^2\right]du^2-dv^2-(dx^3)^2.
\]

\noindent
{\bf 
Manuscript (Brief Communications, Theoretical and Mathematical Physics) received in the editorial office on: 
23.06.88.}

\end{document}